\documentclass[twocolumn,superscriptaddress,aps,prb,showpacs]{revtex4-1}
\usepackage{graphicx,amsmath,amssymb}
\usepackage{epstopdf}
\usepackage{amssymb}
\usepackage{grffile}
\usepackage[usenames]{color}
\usepackage{indentfirst}
\usepackage{float}
\usepackage{color}
\usepackage{mathrsfs}
\usepackage{dcolumn}
\usepackage{bm}
\usepackage[colorlinks=true,bookmarks=false,citecolor=blue,linkcolor=red,urlcolor=blue]{hyperref}

\begin{document}
\title{Thermoelectric Hall conductivity of the fractional quantum Hall systems on a disk}
\author{Zi-Yi Fang}
\affiliation{Department of Physics, Chongqing University,Chongqing 401331, P. R. China}
\author{Dan Ye}
\affiliation{Department of Physics, Chongqing University,Chongqing 401331, P. R. China}
\author{Yu-Yu Zhang}
\email{yuyuzh@cqu.edu.cn}
\affiliation{Department of Physics, Chongqing University,Chongqing 401331, P. R. China}
\author{Zi-Xiang Hu}
\email{zxhu@cqu.edu.cn}
\affiliation{Department of Physics, Chongqing University,Chongqing 401331, P. R. China}
\affiliation{Center for Quantum materials and device, Chongqing University,Chongqing 401331, P. R. China}

\begin{abstract}
For the fractional quantum Hall states on a finite disc, we study the thermoelectric transport properties under the influence of an edge and its reconstruction.  In a recent study on a torus [Phys. Rev. B 101, 241101 (2020)], Sheng and Fu found a universal non-Fermi liquid power-law scaling of the thermoelectric conductivity $\alpha_{xy} \propto T^{\eta}$ for the gapless composite Fermi-liquid state. The exponent $\eta \sim 0.5$ appears an independence of the filling factors and the details of the interactions. In the presence of an edge, we find the properties of the edge spectrum dominants the low-temperature behaviors and breaks the universal scaling law of the thermoelectric conductivity. In order to consider individually the effects of the edge states, the entanglement spectrum in real space is employed and tuned by varying the area of subsystem. In non-Abelian Moore-Read state, the Majorana neutral edge mode is found to have more significant effect than that of the charge mode in the low temperature. 
\end{abstract}

\date{\today }

\maketitle

\section{Introduction}
When the strongly interacting two-dimensional electron gas (2DEG) is exposed in a perpendicular strong magnetic field and extremely low temperature,~\cite{Tsui} the formed fractional quantum Hall (FQH) droplet reveals the non-trivial topological properties, such as the fractional charge excitations and fractional statistics,~\cite{Laughlin83} especially the FQH state~\cite{NayakRMP} at some special filling which could have Majorana zero mode excitation and non-Abelian statistics. In FQH droplet, the bulk is a gapped insulator and the low-lying excitation in the bulk is the neutral magneto-roton excitation~\cite{Girvin} which can be described as pairs of particle-hole excitations.~\cite{boPRL12}  Furthermore, in a Hall bar sample which is used to measure the FQH state,  the existence of an edge is usually unavoidable unless using the Corbino geometry which provides direct access to the longitudinal conductivity in the bulk.~\cite{Corbino, Corbino1} The edge physics plays an essential role to uncover the bulk topological properties due to the bulk-edge correspondence. On a FQH edge, the description of Fermi-liquid theory is broken down and the interacting electrons prefer to form a chiral Luttinger liquid (CLL).~\cite{WenIJPM}  It is theoretically predicted that the current-voltage dependence in the tunneling between a Fermi liquid and a quantum Hall edge obeys a universal power-law $I \sim V^{\alpha}$ where $\nu = 1/\alpha$ is the filling factor.~\cite{Wen94} Such universality, however, has not been conclusively observed in experiments. A very likely reason is the occurrence of additional non-chiral edges that are not tied to the bulk topology,~\cite{ChangRMP03} namely the edge reconstruction, which is a consequence of the competition between the positive background confinement potential and electron-electron Coulomb repulsive interaction.~\cite{Wan02, Wan03, Jain09, Jain10}  It breaks the chirality and universality in the $I-V$ power law. In spite of its non-universality, one of us recently found that in the entanglement spectrum, which represents the virtual edge excitation of the reduced density matrix in a bipartite FQH model wave function, still can be tuned and reconstructed by varying the area of the subsystem.~\cite{Wei}

Since the edge excitation is gapless and has lower energy than that of the neutral magneto-roton excitation in the bulk, intuitively, it dominates the low-lying energy behavior of the system, regardless of the charge or heat transport experiments. Moreover, while backscattering between two nearest neighboring charged edge currents occurs or in the non-Abelian FQH edge, a possible Majorana neutral mode appears. Because of the charge neutrality, this type of edge mode can not be detected directly by electrical measurements. Then the thermoelectric phenomena that provide the direct conversion between heat and electricity are interesting and useful.  In a thermoelectric measurement, one sets up a temperature gradient $\triangledown T$, and an electrical current $I$ is generated by the system to compensate for its effect. They are related to each other by $I_i = -\alpha_{ij} \triangledown_j T$ where $\alpha_{ij}$ is the thermoelectric conductivity. Experimentally, one usually measure the thermopower $S_{xx}$ and Nernst coefficient $S_{xy}$ and they are related to $\alpha_{ij}$ by $S_{ik} = \alpha_{ij}\rho_{jk}$, in which $\rho$ is resistivity. It is known that the thermoelectric response is directly related to thermal entropy per electron.~\cite{Obraztsov, Girvin1982, Bergman2009, Kozii}  Because of the huge degeneracy of the non-Abelian FQH states with quasiparticle excitations, it was proposed to probe the non-Abelian statistics of  the quasiparticles and measure their quantum dimension.~\cite{kun2009,Stern18} Very recently, it is shown that  two-dimensional quantum Hall systems  can reach thermoelectric figure of merit on the order of unity down to low temperature, as a consequence of the thermal entropy from the massive Landau level (LL) degeneracy without considering the Coulomb interaction.~\cite{Fu} 
 
 For a Fermi-liquid, since the entropy is associated with the number of thermal excitations within $k_BT$ near the Fermi energy, a linear $T$ dependence has been conjectured.~\cite{Cooper} While including the Coulomb interaction, the LL degeneracy is lifted, and FQH states are formed at specific fillings factors. In that case, the electron system is a non-Fermi liquid and the ground state has only the degeneracy due to the center of mass translational symmetry in a translational invariant system. In Ref.\onlinecite{Sheng20}, Sheng and Fu calculated the $\alpha_{xy}$ for FQH system in a torus geometry. A non-Fermi liquid pow-law scaling $\alpha_{xy} \propto T^{\eta}$ with $\eta  \sim 0.5$ for composite Fermi-liquid states at $\nu = 1/2$ and $\nu = 1/4$ was found. $\alpha_{xy}$ vanishes exponentially $\alpha_{xy} \sim \exp(-\Delta/k_BT)$ with a neutral magneto-roton gap for $\nu = 1/3$ Laughlin FQH state while $T \rightarrow 0$. However, as we stated above, the low-lying energies of the FQH system are dominated by the edge states in system with a boundary which was neglected in the torus geometry. They should naturally have contributions in the low temperature transport. In this work, we consider the thermoelectric conductivity in disk geometry to see how does the edge state and its reconstruction affect the the thermoelectric conductivity, especially the low temperature scaling behavior. In order to isolate the effect of the edge excitations, the real space entanglement spectrum is used to calculate the $\alpha_{xy}$ for virtual edge excitations in reduced density matrix after tracing the degrees of freedom in subsystem. We will show that the edge, and its reconstruction brings non-universal scaling in low temperature, but still satisfies nonFermi-liquid behavior. The Fermi-liquid behavior for dipolar fermions and importance of the Majorana mode in non-Abelian FQH state are also found.
 
 The rest of this paper is organized as following: In Sec. II, we compare the thermoelectric conductivity results from energy spectrum with and without edge states. The dipolar fermions in FQH and trivial phases are also considered.  In Sec. III, the real space entanglement spectrum is used to analyzing the effect of the edge states. The non-Abelian FQH state and the dependence of $\alpha_{xy}$ are also considered and Sec. IV gives the conclusions and discussions.

\section{comparing different geometries and phases}
Since the thermoelectric conductivity is directly related to the thermal entropy per electron, we perform a fully diagonalizing the Hamiltonian for a finite system at a given filling factor and the energy spectrum are used to calculate the partition function $Z$ as a function of temperature $T$ 
 \begin{equation}
  Z = \sum\limits_{i}\exp(-E_i/k_BT).
\end{equation} 
Then the thermal entropy can be obtained as 
\begin{equation}
  S = -\sum\limits_{i}\rho_i\ln{\rho_i}
\end{equation} 
with $E_i$ the eigenvalue and $\rho_i = \exp(-E_i/k_BT)/Z$ the probability for the $i$-th state to be occupied at temperature $T$. In the absence of disorder, thermoelectric Hall conductivity $\alpha_{xy}$ is proportional to the entropy density as 
\begin{equation}
	\alpha_{xy} = \frac{s}{B} = \frac{S}{N_{orb}}
\end{equation}
 where $N_{orb}$ is the number of orbitals for electrons occupying in Landau level and $s$ is $S$ divided by the area of 2DEG which is $2\pi l_B^2N_{orb}$. Here $l_B$ is the magnetic length $\l_B = \sqrt{\hbar c/eB}$. Suppose the dimension of the Hilbert space is $\Omega$, in high temperature limit such that the thermal activation overwhelms all the energy scale of the many-body system, such as the neutral gap or quasiparticle excitation gap in the bulk, namely $k_BT \gg \Delta$. Each  eigenstate has identical probability $\rho_i = 1/\Omega$ and thus the thermal entropy is classical $S = \ln\Omega$. In this case, $\alpha_{xy} = \ln\Omega/N_{orb}$ is a constant. For a state with $N_e$ electrons at filling $\nu = N_e / N_{orb}$, the dimension of the Hilbert space $\Omega = C_{N_{orb}}^{N_e}$ reaches maximum at $\nu = 1/2$ and thus the half filling has the largest $\alpha_{xy}$. It has potential to enable thermoelectric cooling and power generation with unprecedented efficiency.~\cite{Fu}

\begin{figure}
\includegraphics[width=8.5cm,height=6.5cm]{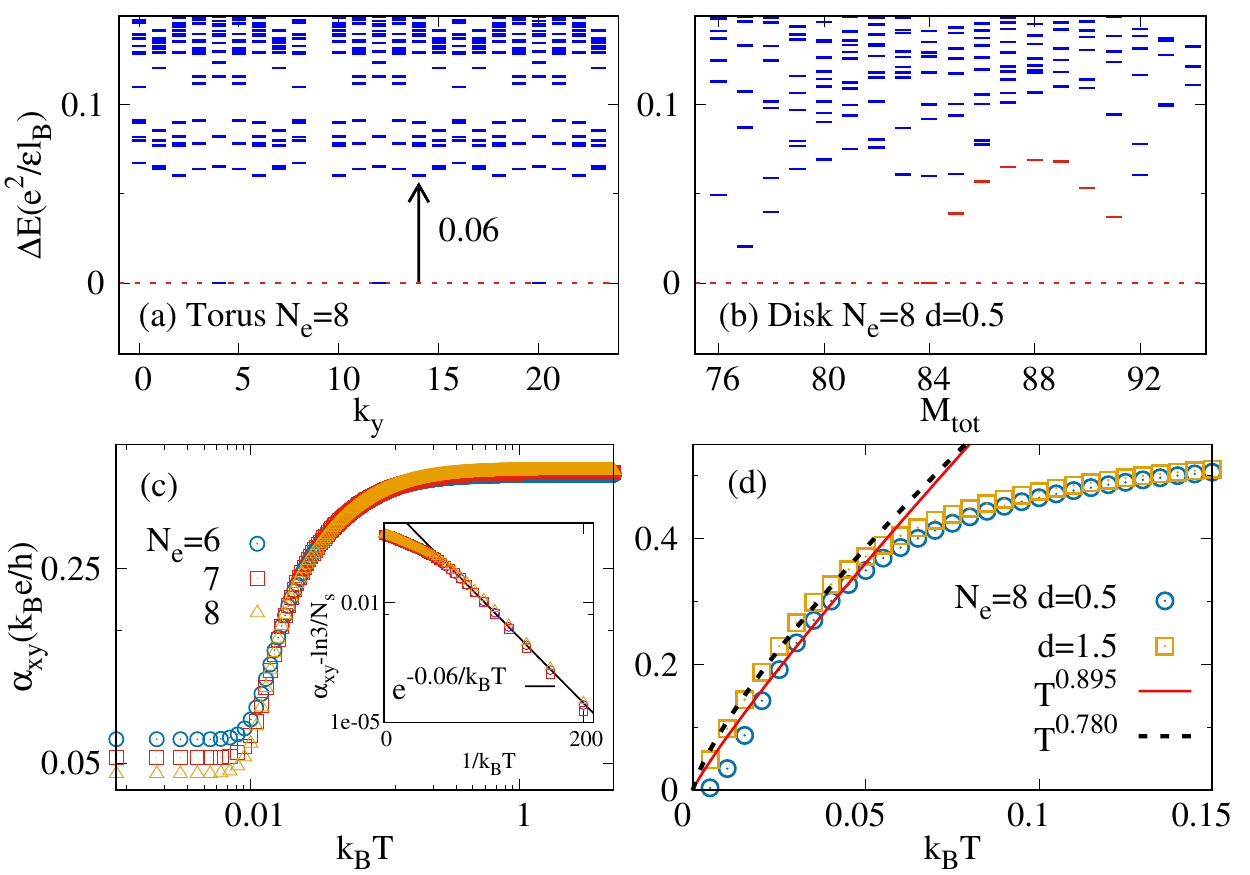}
\caption{\label{fig1} The energy spectrum for 8 electrons in 24 orbitals with Coulomb interaction on torus(a) and disk(b). In fig.(b), the positive confinement is set at a distance $d = 0.5\ell_B$ from electron layer. The gapless edge mode dispersion is labelled by red color. (c) and (d) depict the $\alpha_{xy}$ (in units of $k_Be/h$) versus the thermal energy $k_BT$ in units of $e^2/\epsilon\ell_B =1$ for $6-8$ electrons at $\nu=1/3$. In the inserted plots in (c), we fit the low temperature data with $\alpha_{xy} \sim \exp(-\Delta/k_BT)$. In (d), two confinement potentials are used. One $d = 0.5$ is in the Laughlin phase and the other $d = 1.5$ is after the edge reconstruction. $\eta$ are different for two cases.
}
\end{figure}

At low temperature, as shown in Fig.~\ref{fig1}, we compare the results on a compacted torus with that on a disc. For Coulomb interaction at $\nu = 1/3$, the energy spectrum on torus has three-fold degeneracy due to the center of mass translational symmetry which demonstrates the Laughlin-like state. In this case, the three-fold degenerated ground states are gapped by neutral magneto-roton excitation in the bulk. In disk geometry, mimicking the 2DEG in semiconductor hetero-structure, we diagonalize a Hamiltonian which contains the 2DEG and a homogeneous positive background confinement potential at distance $d$.~\cite{Wan02,Wan03} As shown in Fig.~\ref{fig1}(b), the ground state is unique and a chiral gapless edge excitation (its dispersion is labelled in red color) dominates the low-lying energy spectrum in thermodynamic limit. As a result, the thermoelectric Hall conductivity $\alpha_{xy}$ at low temperature exhibits different behaviors in two cases. In Fig.~\ref{fig1}(c) which was also shown in Ref.~\onlinecite{Sheng20}, the $\alpha_{xy}$ is saturated at a value $\ln3/N_{orb}$ while $T \rightarrow 0$ due to the three-fold ground state degeneracy.  For a gapped phase on torus, the low-temperature data before saturation could be fitted by $\exp(-\Delta/k_BT)$ as shown in the inserted plot. $\Delta$ is actually the neutral gap which is labelled by arrow in Fig.~\ref{fig1}(a). In disk geometry, as shown in Fig.~\ref{fig1}(d), the $\alpha_{xy}$ for $\nu=1/3$ FQH state has a similar behavior as that for the composite Fermi-liquid at $\nu = 1/2$ and $\nu = 1/4$ on torus. It approaches to zero while $T \rightarrow 0$. After neglecting the data below the average energy gap in the spectrum, which origins from the finite size effects, we fit the low-temperature data using $\alpha_{xy} \propto T^{\eta}$ with $\eta \simeq 0.895$ for $d = 0.5l_B$ and $\eta \simeq 0.780$ for $d = 1.5l_B$. The reason to take these two cases is that the Laughlin-like ground state which has total angular momentum $M_{tot} = 3N_e(N_e-1)/2$, only survives as the global ground state in window $d \in [0,1.35]l_B$ for $8$ electrons. When $d > 1.35l_B$, the ground state angular momentum is changed and the reconstruction occurs on the edge. In the presence of an edge, we find varying confinement potential significantly changes the low temperature scaling behavior of the $\alpha_{xy}$.

\begin{figure}
	\includegraphics[width=8.5cm]{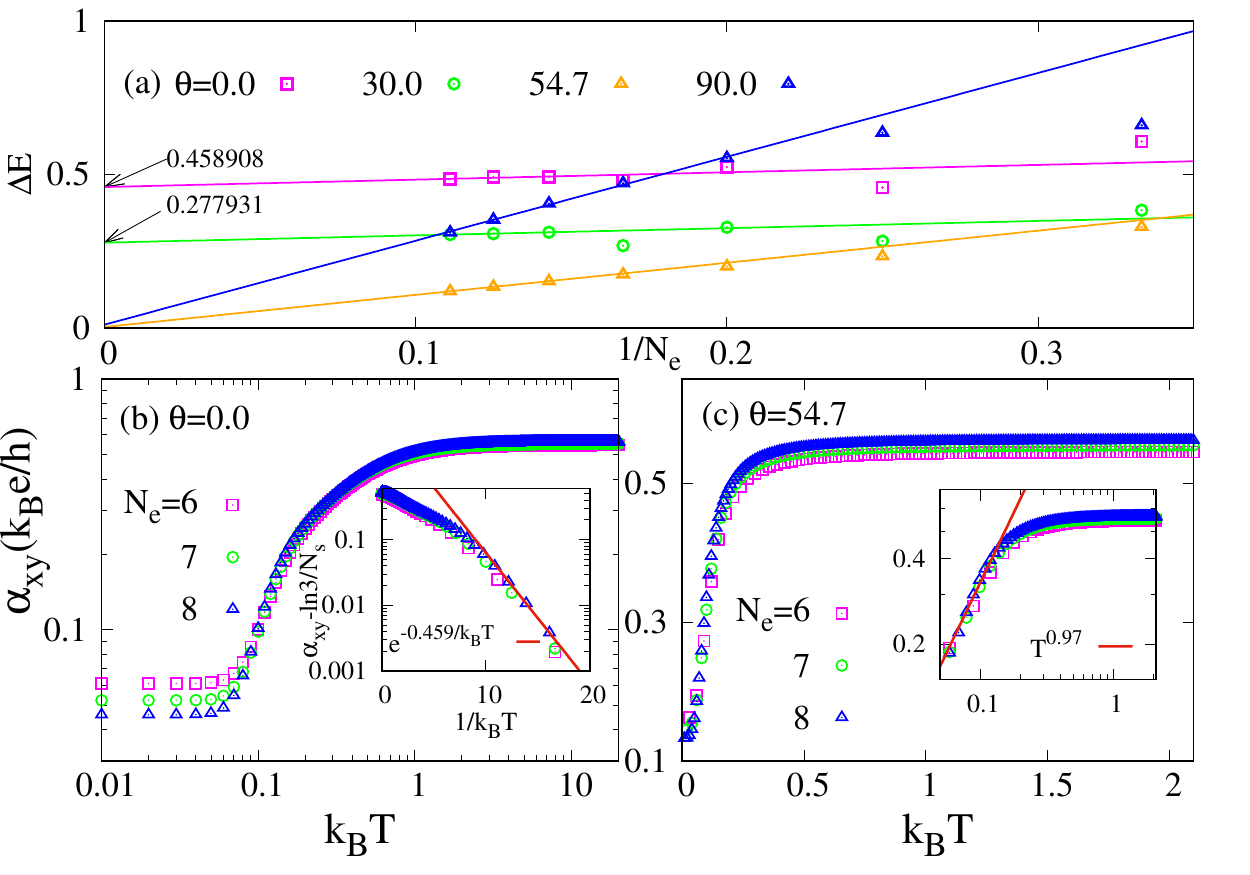}
	\caption{\label{dipoleplot} For fast rotated dipolar fermions on a torus, finite size scalings of the ground energy gap at different titled angles $\theta$ for the dipole moment. The particle number is $8$ and the thickness in $z$ direction is set to $q=0.01$ in unit of $l = \sqrt{\hbar/(2\mu\omega)}$ where $\mu$ is the effective mass and $\omega$ is the frequency of the trap potential.  $\theta = 54.7^\circ$ is the magic angle at which the dipole-dipole interacting is purely anisotropic and the system is gapless. (b) and (c) depict the $\alpha_{xy}$ and fittings for $6-8$ particles at $\theta=0^\circ$ and $\theta=54.7^\circ$. The gapless phase shows the Fermi-liquid behavior with $\eta \simeq 1$.}
\end{figure}

As another example to see the different low temperature behaviors of $\alpha_{xy}$ in gapped and gapless phases, we consider the dipole-dipole interaction for degenerate quantum gas in a fast rotating trap.~\cite{Cooperdipole}
\begin{equation}\label{dipole}
V_{dd}(\vec{r}, \theta) = \frac{r^2-3(z \cos\theta +x \sin\theta)^2}{r^5},
\end{equation}
where $\theta$ is the angle between the dipole moment and the $z$ axis. In a fast rotating limit at which the rotating frequency is close to that of the harmonic trap potential, the system enters into the quantum Hall regime.~\cite{Cooperdipole} It was found~\cite{qiu2011,hu18} that the Laughlin phase robustly survives while the tilting angle $\theta$ is small. As the $\theta$ is increased, the ground state gap is gradually reduced and finally closed, then the system enters into a cluster state~\cite{hu18} because of the anisotropic attractive interaction in one direction. The interaction in Eq.(\ref{dipole}) could be re-arranged as
\begin{equation}\label{dipole1}
V_{dd}(\vec{r}, \theta) = \frac{3\cos^2{\theta}-1}{2} \frac{r^2-3z^2}{r^5} + \frac{3\sin^2{\theta}}{2}\frac{y^2-x^2}{r^5}.
\end{equation} 
The first term is isotropic and the second term is anisotropic in the 2D plane. The magic angle $\theta = 54.7^\circ$ satisfies the condition $3\cos^2{\theta}-1 = 0$, at which the interaction only contains the second term. In Fig.~\ref{dipoleplot}(a), we plot the energy gap between the lowest two states via diagonalizing Eq.(\ref{dipole1}) on torus for several system sizes. The extrapolation in thermodynamic limit from finite size scaling clearly tells us the system is gapped for $\theta = 0^\circ, 30^\circ$ and gapless for $\theta = 54.7^\circ, 90^\circ$ respectively. In Fig.~\ref{dipoleplot}(b) and (c), we plot the $\alpha_{xy}$ for $\theta = 0^\circ$ and  $54.7^\circ$ which are in two different phases. The $\alpha_{xy}$ in FQH gapped phase is similar to that of the Coulomb Hamiltonian in Fig.~\ref{fig1}(a) and the low temperature part still satisfies $\alpha_{xy} \sim \exp(-\Delta/k_BT)$. However, in a topological trivial gapless phase at $1/3$ filling, as shown in In Fig.~\ref{dipoleplot}(c), we find the low temperature behavior could be fitted by $\alpha_{xy} \sim T^{0.97}$ which is very close to the Fermi-liquid exponent $\eta = 1$. Therefore, from low temperature thermoelectric Hall conductivity, we speculate the phase transition in dipolar fermions is a topological phase to Fermi-liquid transition.

Based on the results of above two cases, we suspect that the existence of the gapless edge mode or gapless trivial phase could have different scaling exponent of the thermoelectric Hall conductivity in low temperature.

\section{The effects of edge states}
In previous section, we observe that the presence of an edge significantly changes the low-temperature behaviors of the thermoelectric Hall conductivity. In order to individually consider the edge states, we calculate $\alpha_{xy}$ for the real space entanglement spectrum~\cite{Regnault12,Dubail12, Simon12} of a model wave function.  To be precise, a bipartition of a quantum system is defined when the Hilbert space is divided into two subsystems, $H = H_A\bigotimes H_B$. Then the ground state wave function is performed by a Schmidt decomposition:\cite{Nielsen,Kitaev}
\begin{equation}
   |\Psi\rangle = \sum_i e^{-\frac{1}{2}\xi_i} |\psi_A^i\rangle \bigotimes |\psi_B^i\rangle) 
\end{equation} 
where the $\exp(-\xi_i) = \lambda_i$ are the eigenvalues of the reduced density matrix of subsystem $\rho_A = \text{Tr}_B \rho$. It is normalized by $\sum_i e^{-\xi_i}=|\psi\rangle\langle\psi|=1$.  If $\rho_A$ is finite dimensional, the von Neumann entanglement entropy is defined as $S_A = -\sum_i \lambda_i \text{log} \lambda_i = \sum_i \xi_i \exp(-\xi_i)$. It was known by Haldane~\cite{HaldaneES} that the full structure of the ``entanglement spectrum" (ES) which is the logarithmic Schmidt spectrum of level $\xi_i$ contains much more information about the entanglement between two halves than that from $S_A$ only.  It plays a key role in analyzing the topological order of the FQH state. The structure of ES is analogous to the low energy excitations of a many-body Hamiltonian. Especially, for the model wave function, the counting per momentum sector of ES is identical to the energy spectrum for the edge excitations.  Beyond the counting, one of us~\cite{Wei} recently found that the entanglement spectrum in real space cut appears the signals of edge reconstruction via tuning the area of the subsystem. The edge velocity reaches its minimum when the area of subsystem is equal to that of the correlation hole.

\begin{figure}
\centering
\includegraphics[width=9cm]{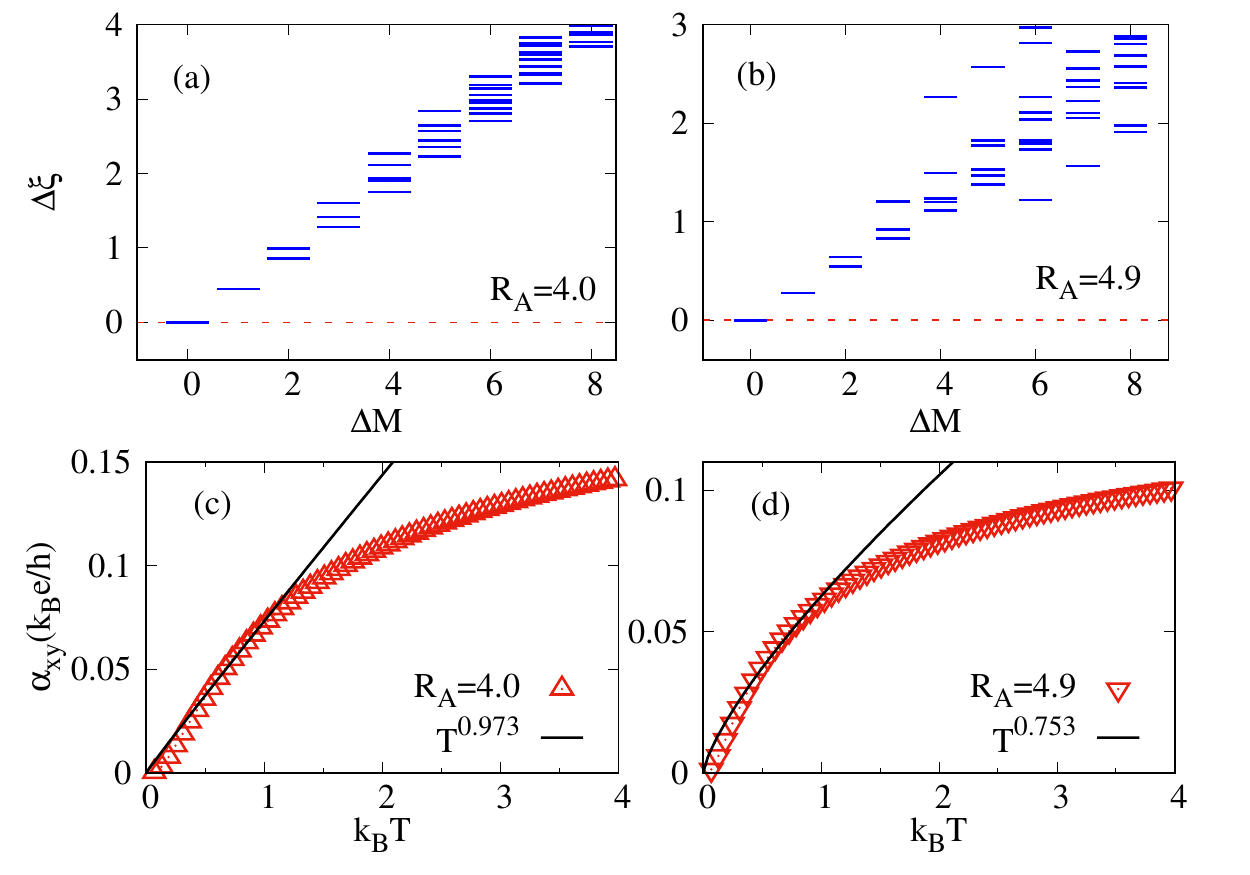}
\caption{\label{fig3} The real space ES for subsystem of $N_A = 4$,  $R_A=4.0l_B$(a) and $R_A=4.9l_B$(b) in a $10$-electron Laughlin state. $N_A$ and $R_A$ is number of electrons and radius of disk for the subsystem. The typical energy spacing is defined as the gap between the lowest two states. (c) and (d) depict the $\alpha_{xy}$ and low temperature fittings from the spectrum (a) and (b) correspondingly. Their fitting exponents are $\eta \simeq 0.973$ and $\eta \simeq 0.753$ respectively.}
\end{figure}

We consider the $\nu=1/3$ Laughlin wave function for $10$ electrons on a disk which can be obtained either from diagonalizing a $V_1$ pseudopotential hamiltonian~\cite{HaldanePP} or from the Jacks  polynomial recipe.~\cite{Bernevig1, Bernevig2} Because the bipartition we used conserves the rotational symmetry, the ES for a given number of electrons in subsystem $N_A$ and radius of the circular cut $R_A$ are shown in Fig.~\ref{fig3}(a) and (b). Here we consider the subsystem contains $4$ particles and the radius of the subsystem $R_A$ is a parameter we tuned. It shows that the ES for the same $N_A$ and different $R_A$ have different spectra. Consequently, their thermoelectric Hall coefficient $\alpha_{xy}$ have different behaviors at low temperature as shown in Fig.~\ref{fig3}(c) and (d) respectively. To eliminating the finite size effect, we still fit the low temperature data above the typical average energy interval. The exponent are $\eta \simeq 0.973$ and $\eta \simeq 0.753$ for two cases, both of which are larger than $\eta \simeq 0.5$ on torus and smaller than that of the Fermi-liquid value. Therefore, the FQH edge states still possess non-Fermi-liquid characteristics, but its scaling exponent could be changed by tuning the edge velocity (slope of the dispersion) and thus non-universal.

\begin{figure}[htbp]
\centering
\includegraphics[width=8.5cm]{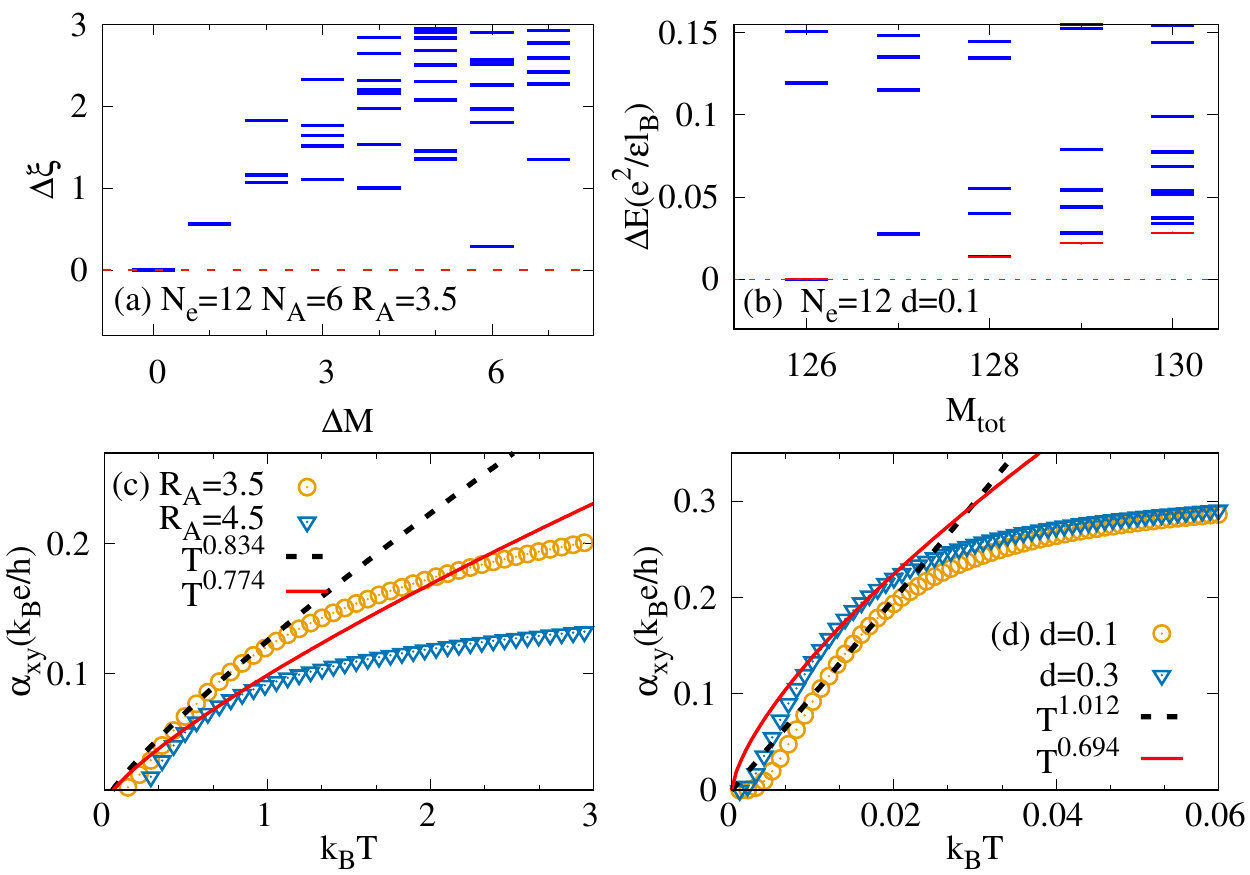}
\caption{\label{fig4} (a) ES for the $\nu=5/2$ Moore-Read state $N_e = 12$, $N_A=6$,$R_A= 3.5$. (b) The low-lying energy spectrum of a mixed Hamiltonian with $\lambda = 0.5$ for $12$ electrons in different confinement potentials. The red bars are the Majorana neutral edge mode which has lower energy than that of the bosonic charge mode. (c) and (d) depicts the $\alpha_{xy}$ versus $k_B T$ for (a) and (b) correspondingly. In each plots, we take two $R_A$s and $d$s for the sake of observing the different fittings in different edge spectrum.
}
\end{figure}

Now let's move to the non-Abelian FQH liquid which has more complicate edge structures. One prominent example is the Moore-Read state~\cite{Moore} which is the model wave function for FQH state at $\nu = 5/2$ filling in the first Landau level. The model wave function could also be obtained from diagonalizing a three-body $V_3$ pseudopotential~\cite{Simon07} Hamiltonian or Jacks. Fig.~\ref{fig4}(a) shows the ES of a $12$-electron Moore-Read wave function. Unlike the ES for Laughlin state, the counting of edge spectrum becomes to be $1,1,3,5\cdots$. This is due to the existence of a neutral Majorana mode besides the bosonic charge mode.~\cite{Wan08} Fig.~\ref{fig4}(b) depicts the energy spectrum of a mixed Hamiltonian 
\begin{equation}
	H = (\lambda - 1) H_{2B} + \lambda H_{3B}
\end{equation}
with $\lambda = 0.5$ for 12 electrons under different strength of the confinement potentials. $H_{2B}$ contains the electron-electron Coulomb interaction and confinement potential. $H_{3B}$ is the three-body pseudopotential Hamiltonian. The reason of mixing $H_{3B}$ is to emphasize the edge states which have zero energy in the model Hamiltonian. In this case, the low-lying edge states are separated from the bulk states and they clearly show the same counting rule as that in (a). The Majorana edge states are labelled by red bars. Fig.~\ref{fig4}(c) and (d) are their corresponding $\alpha_{xy}$ and low temperature fittings. For the ES, we have $\eta \simeq 0.834$ and $\eta \simeq 0.774$ for $R_A = 3.5$ and $4.5$ respectively. Again, tuning the area of the subsystem or the edge velocities changes the scaling exponent $\eta$. For the energy spectrum which is a mixture of edge and bulk states, we obtain different results for different confinement potentials. In the disk geometry, we found~\cite{Wan08} the Moore-Read state as the global ground state exists in range $d \in [0,0.5]l_B$. In Fig.~\ref{fig4}(d) we plot the $\alpha_{xy}$ for $d = 0.1l_B$ and $d = 0.3l_B$ the fitting exponents are $\eta \simeq 1.012$ and $\eta \simeq 0.694$ respectively. In all cases, the $\eta$ decreases as decreasing the strength of the confinement potential (increasing $d$ in energy spectrum or increasing $R_A$ in ES before reconstruction). On the other hand, it is known~\cite{Hu09,Hu16} that the neutral edge mode has one order lower velocity or excitation energy than that of charge mode. From the $\alpha_{xy}$ as shown in Fig.~\ref{fig4}(d), the low temperature behaviors are dominated near the scale $k_BT\sim 0.01$ which has the same order of the energy difference between the ground state and the lowest edge state in neutral edge mode. Therefore, for the non-Abelian FQH state, the Majorana neutral edge mode has much more significant contribution in thermoelectric conductivity and could be detected in the low temperature thermal Hall experiments.~\cite{kun2009, Stern18}
 
\begin{figure}[htbp]
\centering
\includegraphics[width=8cm,height=4cm]{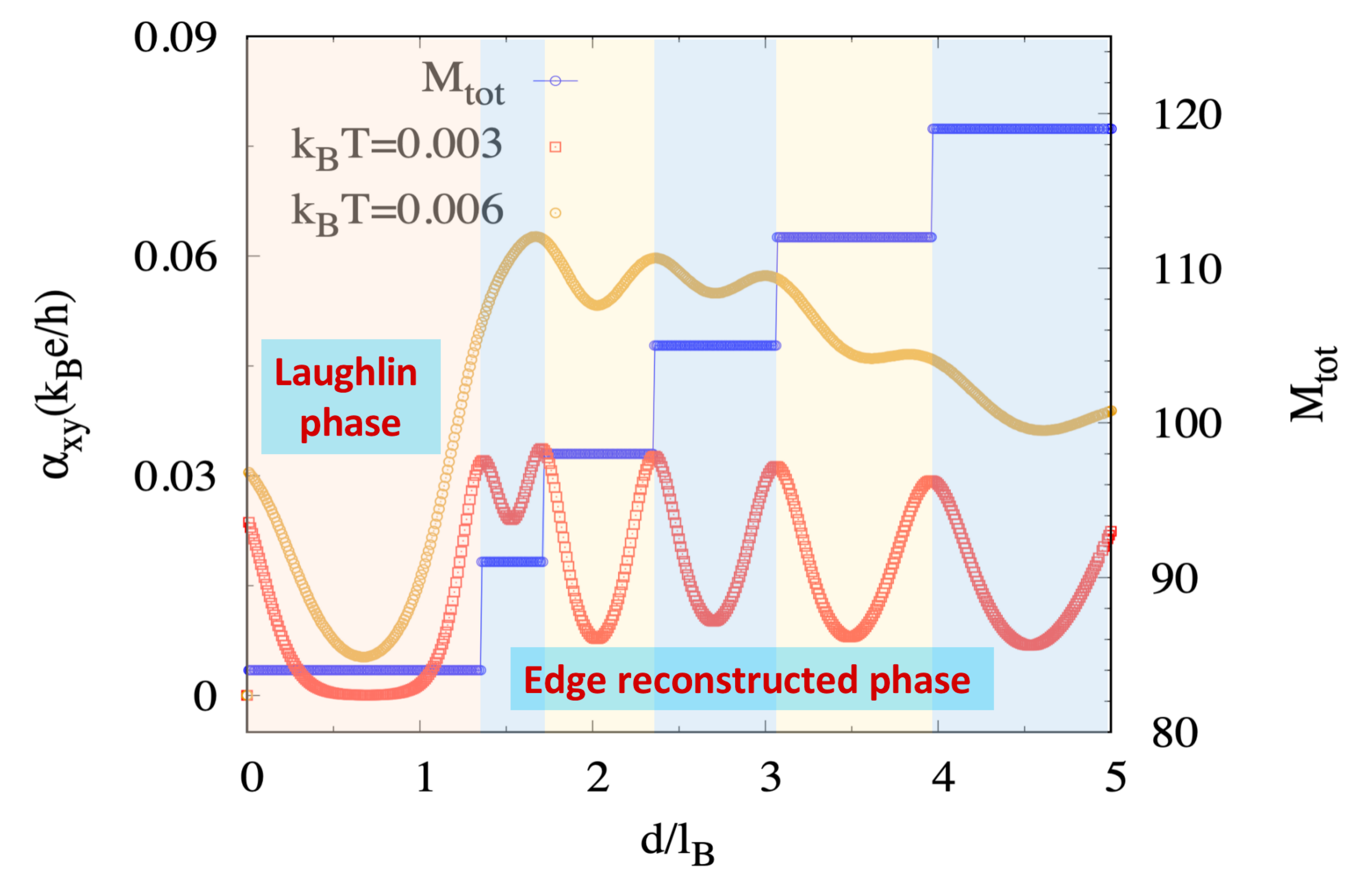}
\caption{\label{fig5} The ground state quantum number $M_{tot}$ and low temperature $\alpha_{xy}$ versus the distance between 2DEG and confinement potential $d/l_B$ for $8$ electrons at $\nu=1/3$. The peak of $\alpha_{xy}$ corresponds to the ground state phase transition, or the edge reconstruction.}
\end{figure}
 
 Besides the scaling behavior in low temperature, the edge confinement also has effects on the value of $\alpha_{xy}$ itself. In Fig.~\ref{fig5}, we plot the ground state angular momentum and $\alpha_{xy}$ as a function of the confinement parameter $d/l_B$. Two temperatures are considered. The $M_{tot}$ is the total angular momentum of the ground state which is $3N_e(N_e-1)/2$ for Laughlin state. While increasing $d$, the electrons firstly stay stably in FQH phase and then form stripes at the edge which is accompanied with a jump of $M_{tot}$.~\cite{WenIJPM,Wan02}. If we treat the $M_{tot}$ as the ground state quantum number, the jumps in  $M_{tot}$ correspond to ground state phase transitions and each $M_{tot}$ could be treated as a phase. It is interesting to see that in the Laughlin phase for $d < 1.35l_B$, thermoelectric Hall conductivity $\alpha_{xy}$ is smaller than that of the reconstructed phase. This is because the edge reconstruction $d > 1.35l_B$ induces more edge modes and thus make more contributions to the thermal transport. Moreover, the $\alpha_{xy}$ has maximum at the phase transition and minimum at the center of each phases. As the temperature increasing, the structure of $\alpha_{xy}$ is gradually erased by thermal excitations.
 
\section{Discussions and Conclusions}
In this work, we study the thermoelectric Hall conductivity of the FQH system in disk geometry, especially its low temperature behaviors. Because of the presence of gapless edge mode, the low temperature behaviors of $\alpha_{xy}$ in disk geometry are similar to that of the gapless composite Fermi-liquid state on a torus. Because of the edge spectrum could be changed by tuning the strength of the background confinement in energy spectrum or the area of subsystem in the bi-partition ES, the scaling exponent $\eta$ of $\alpha_{xy} \sim T^{\eta}$ in low temperature has strong dependence of the system parameters. Comparing to the results on torus, we find $\eta$ decreases as decreasing the strength of the confinement potential (increasing $d$ in energy spectrum or increasing $R_A$ in ES before reconstruction). For the dipolar fermions at the magic angle, we verify the Fermi-liquid scaling behavior with $\eta \sim 1.0$ in the gapless phase. For non-Abelian Moore-Read state at $\nu = 5/2$, we find the Majorana neutral edge mode has more significant contribution to the thermoelectric Hall conductivity and dominates the low temperature scaling behavior, explaining the reason of the non-Abelian signals appear in the low temperature thermal Hall experiments. 

In conclusion, comparing to the results on torus, we find the edge excitations of the FQH states dominate the low temperature scaling behavior of $\alpha_{xy} \sim T^{\eta}$. Both the $\alpha_{xy}$ and its scaling exponent $\eta$ are strongly affected by the edge confinement which makes the scaling non-universal. More edge modes produced by edge reconstruction or the neutral Majorana mode in non-Abelian FQH state contributes significantly to $\alpha_{xy}$ at low temperature thermal Hall transport.

\acknowledgments
This work was supported by National Natural Science Foundation of China
(Grants No.12075040 and No.11974064),  the Chongqing Research Program of Basic Research and Frontier Technology (Grants No.cstc2020jcyj-msxmX0890) and the Fundamental Research Funds for the Central Universities under Grant No. 2020CDJQY-Z003.

\end{document}